\documentclass[a4paper]{jpconf}

\usepackage{graphicx}
\usepackage{cite}

\begin{document}
\title{Relaxation in the 3D ordered CoTAC spin chain by quantum nucleation of 0D domain walls}

\author{E. Lhotel$^1$, E. N. Khatsko$^2$ and C. Paulsen$^1$}

\address{$^1$ Institut N\'eel, CNRS \& Universit\'e Joseph Fourier, BP 166, F-38042 Grenoble Cedex 9, France}
\address{$^2$ Institute for Low Temperature Physics and Engineering, National Academy of Sciences of Ukraine,  310 164 Kharkov, Ukraine.}

\ead{elsa.lhotel@grenoble.cnrs.fr}

\begin{abstract}
We have shown that resonant quantum tunnelling of the magnetisation (QTM), until now observed only in 0D cluster systems (SMMs), occurs in the molecular Ising spin chain, CoTAC ([(CH$_3$)$_3$NH]CoCl$_3$ - 2H$_2$O) which orders as a canted 3D-antiferromagnet at $T_C=4.15$ K. This effect was observed around a resonant like field value of 1025 Oe. 
We present here measurements of the relaxation of the magnetisation as a function of time, from the zero field cooled (ZFC) antiferromagnet  state and from the saturated ferromagnet state. We show that, at the resonant field, the relaxation from the saturated state occurs in a complicated process, whereas, surprisingly, in the case of the ZFC state, the relaxation is
exponential. 
\end{abstract}

\begin{figure}[h]
\begin{center}
\includegraphics[height=3.2cm]{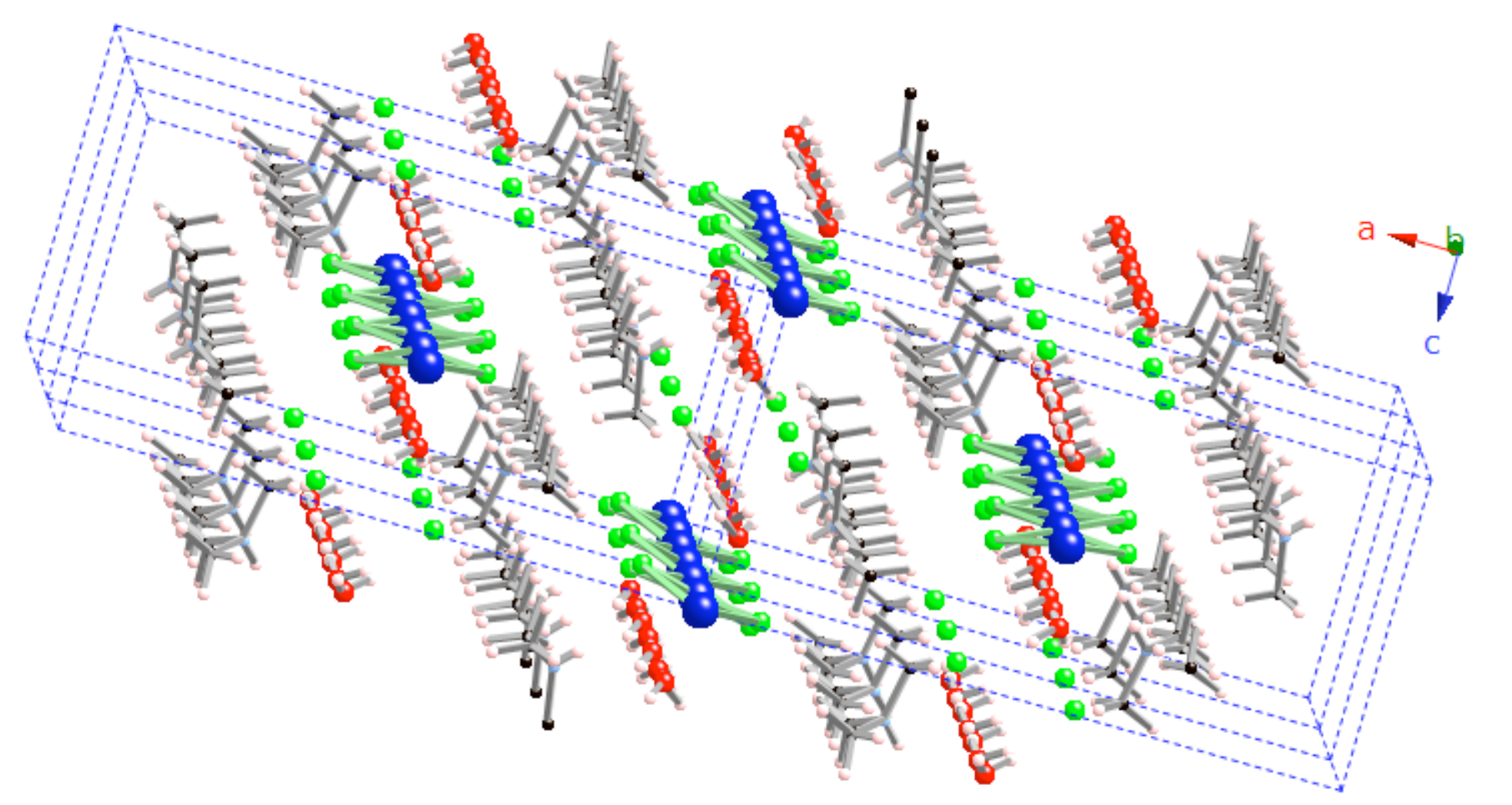} \qquad \qquad
\includegraphics[height=3.2cm]{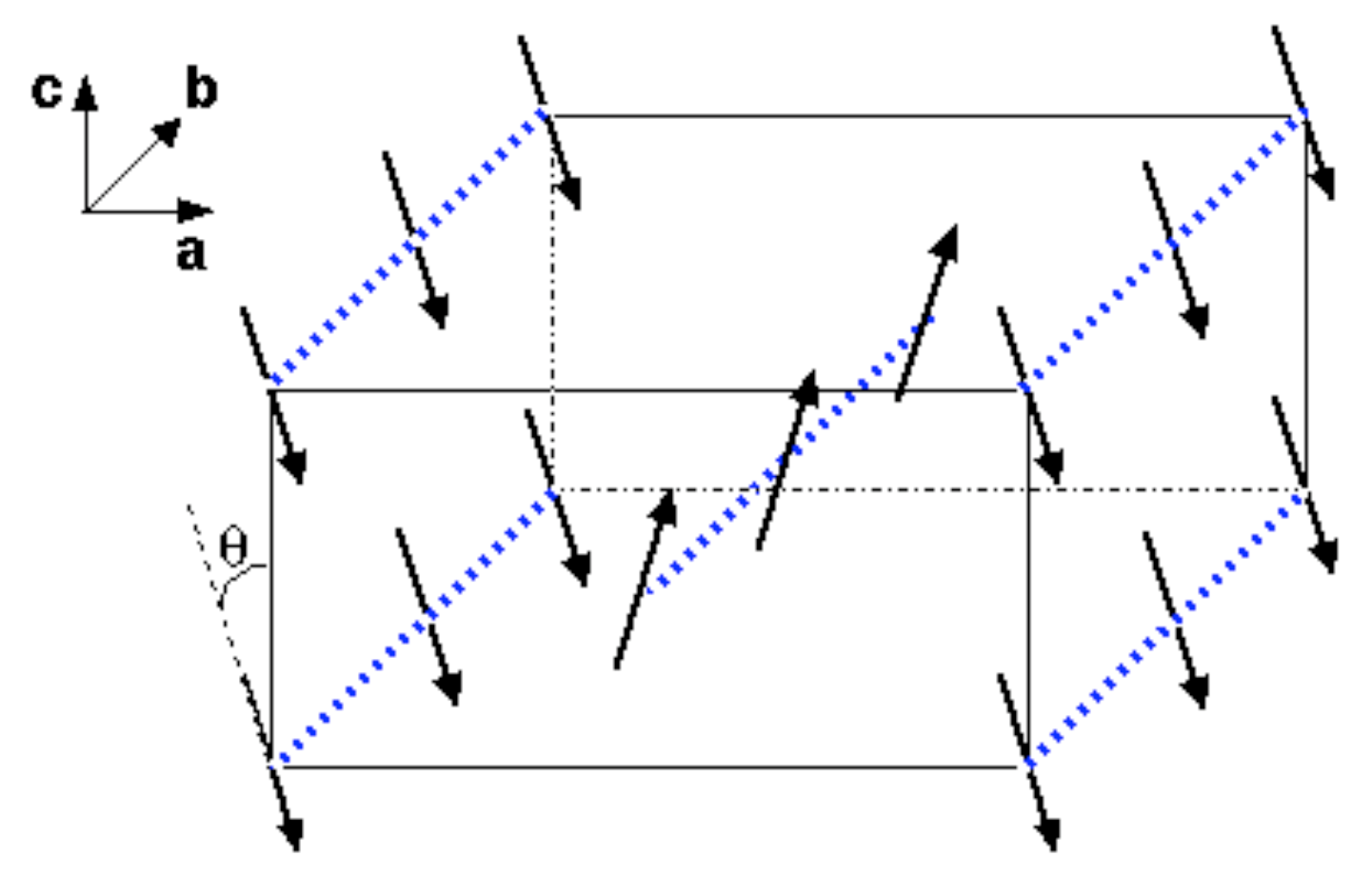}
\caption{\label{CoTAC} Left: Representation of the crystallographic structure of CoTAC with Co (blue), Cl (green), O (red), H (pink), C and N (black)atoms. Right: Representation of the magnetic structure. The dashed lines represent the chains direction. The canting angle $\theta$ is equal to 10$^{\circ}$.}
\end{center}
\end{figure}

Synthetic chemistry has succeeded in designing a wide range of molecular magnets which exhibit original magnetic properties. Single molecule magnets (SMMs), which are zero-dimensional systems (0D), are made of well isolated clusters with a well-defined macro-spin S, a large Ising-like anisotropy, and a weak magnetic coupling with its neighbours. These SMMs have been shown to be model systems for investigating quantum tunnelling of the magnetisation (QTM)  \cite{Sessoli93, Sangregorio97}. 
Recently, we have reported the observation of resonant QTM in a three-dimensional (3D) ordered system, [(CH$_3$)$_3$NH]CoCl$_3 \cdot $2H$_2$O (CoTAC) \cite{Lhotel06}. CoTAC is a 3D network of exchange-coupled ferromagnetic chains which orders antiferromagnetically below 4.15 K \cite{Losee73, Bruckel93, Kobets02, Groenendijk82} (See Figure \ref{CoTAC}). The magnetic structure is shown in the right hand side of Figure \ref{CoTAC}: The ferromagnetic Co chains run along the $b$ axis. A weak ferromagnetism develops along the $a$-axis. The main component of the magnetisation is parallel to the $c$-axis (which is the easy magnetisation axis) and is antiferromagnetic from one chain to the other in the $a$ direction. 
Due to the presence of  3D long-range order, this system is essentially different from SMMs because there is no well isolated entity, such as a macro spin, which can tunnel. We proposed that the observed phenomena are caused by the nucleation of 0D domain-walls along the chains. Based on exchange couplings and anisotropy measurements, we first showed that the domain walls are made up of a very small number (8 or so) of spins. We proposed that the resonant effects are due to the level crossing between a state containing one of these walls, and one which has not. We could then show with a crude model that the Zeeman energy at the resonant field is of the order of the energy of a domain-wall nucleation. The energy barrier is caused by  the anisotropy and exchange coupling between the chains.

In this paper, we present measurements of the relaxation of the magnetisation, with the field applied along the $c$-axis, and at the resonant field (1025 Oe), starting from the saturated state and from the zero field cooled (ZFC) state. We show that the relaxation behavior is essentially different from what is observed in the SMMs and that we can describe the main features by considering the different possible magnetic configurations.  

The measurements were carried out with a SQUID magnetometer developed at the Institut N\'eel equipped with a miniature $^3$He-$^4$He dilution refrigerator allowing measurements down to 70 mK. Absolute values of the magnetisation are made by the extraction method. The sample is a CoTAC single crystal of 12.2 mg. 

\begin{figure}[h]
\begin{center}
\includegraphics[height=6cm]{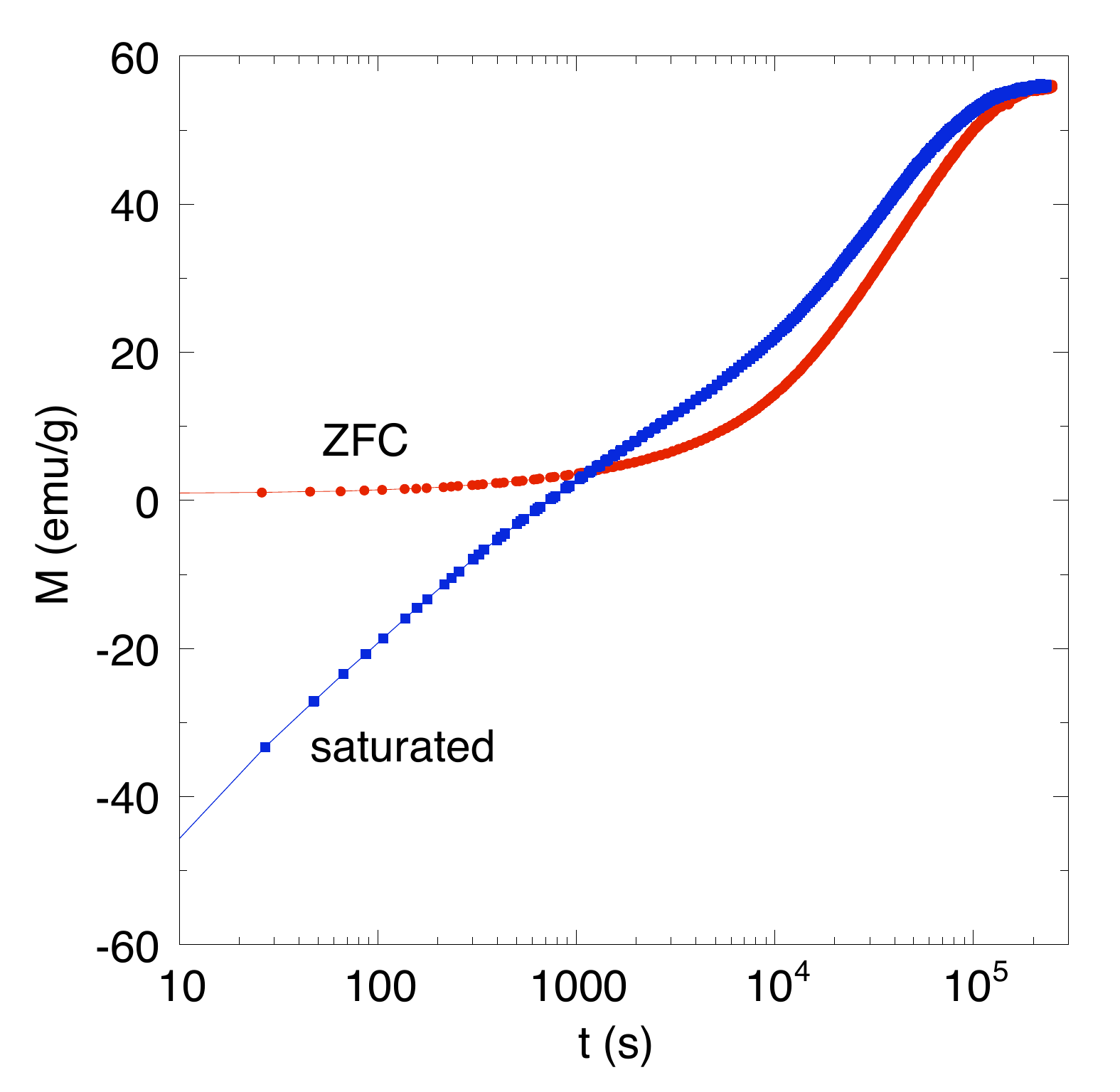}
\hspace{2pc}
\raisebox{4ex}{\begin{minipage}[b]{7cm}
\caption{\label{figrelax} Magnetisation $M$ as a function of time $t$ in a semi-logarithmic scale in a 1025 Oe applied field and at a temperature of 75 mK. Red points are from ZFC state. Blue squares are from the (negative) saturated state.}
\end{minipage}}
\end{center}
\end{figure}

\begin{figure}[h]
\includegraphics[width=16cm]{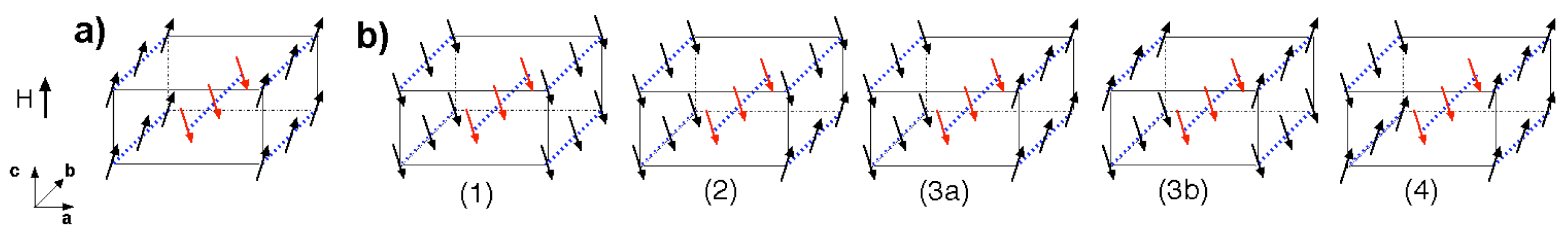}
\caption{\label{schema} Representation of different magnetic configurations of the nearest neighbour chains for a down chain (in the centre- in red) which is going to reverse: a) case of the ZFC antiferromagnetic state. b) five configurations that occur when relaxation is from the saturated ferromagnetic state. }
\end{figure}

Figure \ref{figrelax} shows the relaxation of the magnetisation at 75 mK as a function of time, up to $2 \times 10^5$  ~s (more than 2 days and 7 hours) from two different initial states in a semi-logarithmic plot. In the ZFC state, the sample was first heated above the transition temperature (4.15 K) and then cooled in zero field, before applying a 1025 Oe field. For the (negative) saturated state, a negative - 2000 Oe field was applied to saturate the sample, then the field was changed to the same (positive) 1025 Oe measuring field. 
The first striking result is that the two curves cross for a relaxation time of 1000 s, and so the relaxation from the saturated state is much faster than the relaxation from the ZFC state. This observation indicates that the relaxation process depends on the initial spin configuration. Indeed, when the relaxation starts from the ZFC state, the sample is in an antiferromagnetic configuration (the ground state at H=0) corresponding to the magnetic structure of the left of Figure \ref{CoTAC}  (see also Figure \ref{schema}a). On the contrary, when starting from the saturated state, the structure is ferromagnetic (see also Figure \ref{schema}b.1). As a consequence, the local environment for the nucleation of domain walls along the chains is different for the two cases and will lead to faster or slower nucleation times as we shall explain.

\begin{figure}[h]
\begin{minipage}{7.5cm}
\includegraphics[height=6.4cm]{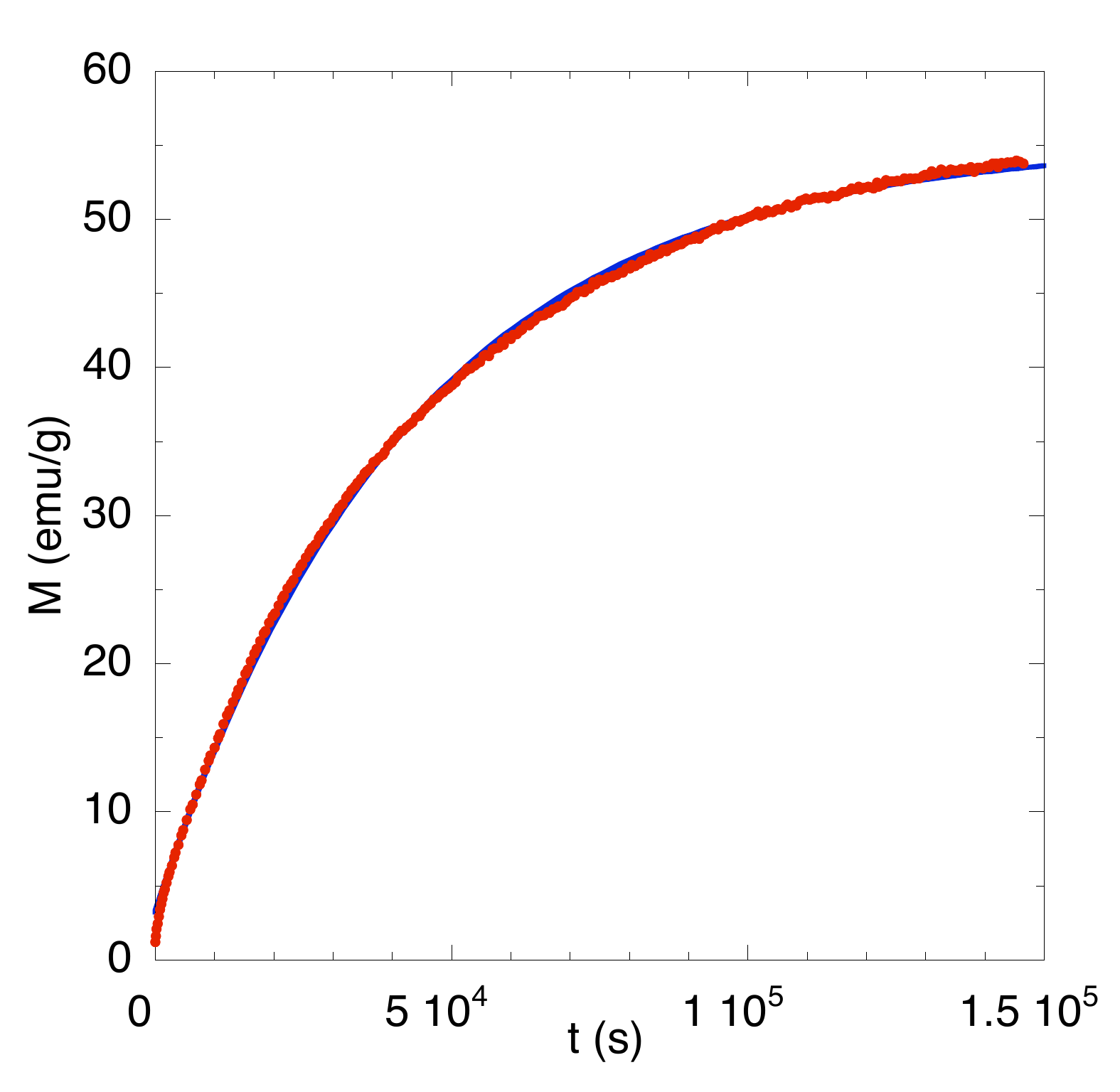}
\end{minipage}
\hspace{1pc}
\begin{minipage}{7.5cm}
\includegraphics[height=6.4cm]{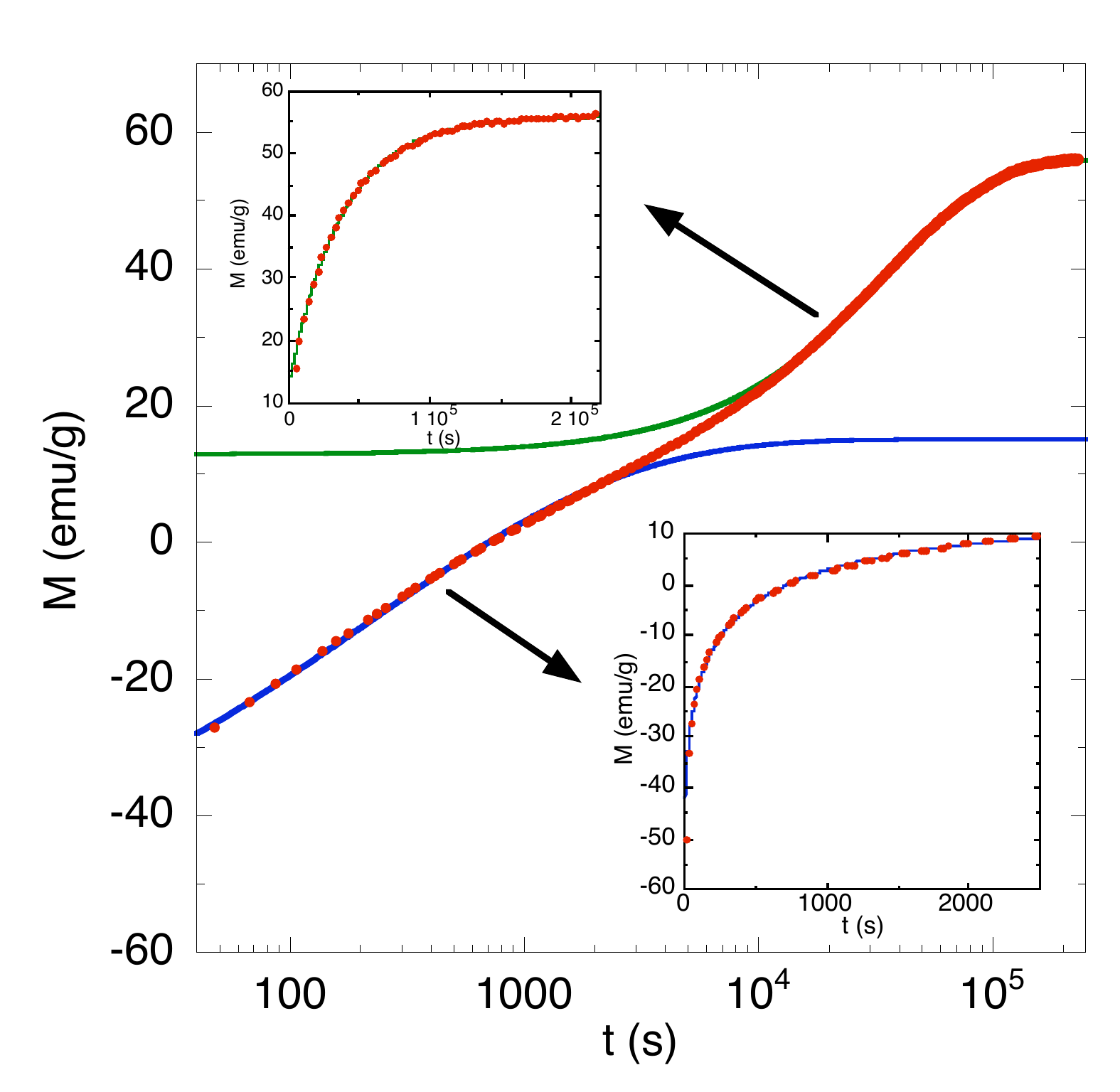}
\end{minipage}
\begin{minipage}{7.5cm}
\caption{\label{figrelaxZFC} Magnetisation $M$ as a function of time $t$, starting from a ZFC state in a 1025 Oe applied field and at a temperature of 75 mK. The line is a fit to an exponential behaviour: $ M=55.1 - 51.9 \exp(-t/\tau)$ with $\tau=4.26 \times 10^4$~s. }
\end{minipage}
\hspace{1pc}
\begin{minipage}{7.5cm}
\caption{\label{figrelaxsat} Magnetisation $M$ as a function of time $t$ in a semi-logarithmic scale, in 1025 Oe at 75 mK, starting from a saturated state. The lines show the fits (see also the insets in a linear scale): i) at short times (right inset), to the stretched exponential behaviour: $M=15-72\exp[(-t/\tau_1)^{0.39}]$ with $\tau_1=220$~s, ii) at long times (left inset) to the exponential behaviour: $M=55.9-43.1\exp(-t/\tau_2)$ with $\tau_2=3.76 \times 10^4$~s. }
\end{minipage}
\end{figure}

Note that both curves can not be described by a square root function at short times as is observed for QTM in the case of SMMs due local field effects \cite{Prokofev98, Ohm98}. In fact, the relaxation curve from the ZFC state can be fit to a  {\it  very good approximation } over the entire  range by an exponential curve (See Figure \ref{figrelaxZFC}): $M=55.1 - 51.9 \exp(-t/\tau)$ with $\tau=4.26 \times 10^4$~s. Curiously,  out of resonance (for $H \not= 1025$ Oe), the relaxation is no longer exponential. 
Concerning the curve from the saturated state, two time scales can be separated to describe the curve (See Figure~ \ref{figrelaxsat}): at short times ($t<2300$~s), the data can be reasonably fit with a stretched exponential law (See the right inset of Figure \ref{figrelaxsat}): $M=15-72\exp[(-t/\tau_1)^\beta]$ with $\tau_1=220$~s and $\beta=0.39$. The existence of the $\beta$ parameter indicates that several relaxation times are present as short times. 
At long times ($t>7500$~s), the data can again be fit with a simple exponential law (See the left inset of Figure \ref{figrelaxsat}): $M=55.9-43.1\exp(-t/\tau_2)$ with $\tau_2=3.76 \times 10^4$~s. The value of $\tau_2$ is of the same order of magnitude as the $\tau$ value obtained in the case of the ZFC relaxation. This suggests that the relaxation process is of the same nature in both cases. 

We  propose a phenomenological model  to describe the  relaxation behaviour based on considerations about the magnetic configurations as depicted in Figure \ref{schema}a and b.
When the resonant field is applied, spin chains can begin to reverse by quantum nucleation of 0D domain-walls as described in Ref. \citeonline{Lhotel06}.  Each nucleation of a domain wall on a given chain is an independent event, governed by some characteristic relaxation time. Furthermore,  when a domain wall occurs, it will rapidly sweep along the chain reversing spins from one end of the sample to the other, (or at least until pinned by a defect or impurity), thereby greatly amplifying the relaxation for each nucleation event. We suggest that the relaxation time for the nucleation of a 0D wall,  in effect by quantum tunnelling through an energy barrier, depends on the local environment of the chain. Indeed, in this sample, the height of the energy barrier is determined not only by the anisotropy (as is the case for SMMs), but also by the sign of the exchange interactions between neighbouring chains: because the interchain interaction is antiferromagnetic, chains that are anti-aligned will have a lower energy, and thus see a higher barrier, whereas chains that are aligned will have a higher energy, and thus be closer to the top of the barrier. Because the relaxation times are exponentially sensitive to barrier height, relaxation will be much faster in the later case.

Consider first the relaxation from the ZFC antiferromagnetic state (See Figure \ref{schema}a). With the resonant field applied in the up direction, {\it  all  down spin chains } (centre) will see the same local configurational environment: i.e. four nearest neighbouring chains in the up direction. The energy barrier is unique, and the relaxation is observed to  exponential  with the measured characteristic time of the order of $4\times 10^4$~s.

Relaxation from the saturated ferromagnetic state is more complicated. The system will pass through the different configurations, (some concurrently) as shown in Figure \ref{schema}b.  At the start, all four neighbouring chains will be down (Figure \ref{schema}b.1) and the relaxation of any given down chain will be very fast. Soon many down chains will have one neighbouring up chain, with three neighbouring down chains. As the system continues to relax, down chains will have two neighbouring up chains and two neighbouring down chains  (Figure \ref{schema}b.3a or 3b) and so on, until finally nearly all "surviving" down chains are surrounded by four neighbouring up chains, (Figure \ref{schema}a): the antiferromagnetic state is recovered and the relaxation is similar to the ZFC curve. Due to the antiferromagnetic interchain exchange interaction, each configuration has a different energy barrier and hence different time scale, with Figure \ref{schema}b.1 the fastest, Figure \ref{schema}b.2 a bit slower etc. until Figure \ref{schema}a the slowest. 
This explains why we can fit the short time scale data with a stretched exponential function, which accounts for several relaxation times. Indeed, the entire range of data can be fit very nicely with 5 exponential functions, with values of  $\tau=100$~s.  to  $\tau=4 \times 10^4$~s, the last in good agreement with the  ZFC relaxation time.


In conclusion, we have shown that relaxation phenomena in presence of resonant QTM in the 3D ordered CoTAC compound is very unusual and brings into play several relaxation times which vary from 100 to more than $10^4$~s, depending on the spin configuration.

\end{document}